\documentclass{amsart}
\usepackage{amsfonts}
\usepackage{amsmath}
\usepackage{amssymb}

\theoremstyle{plain}
\newtheorem{theorem}{Theorem}[section]
\newtheorem{lemma}[theorem]{Lemma}
\newtheorem{proposition}[theorem]{Proposition}
\theoremstyle{definition}

\theoremstyle{remark}

\numberwithin{equation}{section}

\begin{document}
\title{The Marsden-Weinstein reduction structure of integrable dynamical
systems and a generalized exactly solvable quantum superradiance model}
\author{Bogolubov N.N (Jr.)$^{1)}$, Prykarpatsky Y.A$^{2)}$.}
\address{$^{1)}$The V.A. Steklov Mathematical Institute, Russian Academy of
Sciences, Moscow, Russia\\
and \\
the Abdus Salame International Center of Theoretical Physics, Trieste, Italy%
\\
$^{2)}$ Department of Applied Mathematics, University of Agriculture,
Krakow, Poland}
\date{}
\maketitle

\begin{abstract}
An approach to describing nonlinear Lax type integrable dynamical systems of
modern mathematical and theoretical physics, based on the Marsden-Weinstein
reduction method on canonically symplectic manifolds \ with group symmetry,
is proposed. Its natural relationship with the well known
Adler-Kostant-Souriau-Berezin-Kirillov method and the associated R-matrix
approach is analyzed.

A new generalized exactly solvable spatially one-dimensional quantum
superradiance model, describing a charged fermionic medium interacting with
external electromagnetic field, is suggested. The Lax type operator spectral
problem is presented, the related $R$-structure is calculated. The Hamilton
operator renormalization procedure subject to a physically stable vacuum is
described, the quantum excitations and quantum solitons, related with the
thermodynamical equilibrity of the model, are discussed.
\end{abstract}

\section{Introduction}

As it is well known \cite{Ar,AM,RS-T1,FT}, the most popular canonically
symplectic manifolds are supplied by cotangent spaces $M:=T^{\ast }(P)$ to
some "coordinates"\ phase spaces $P$, which can often possess additional
symmetry properties. If this symmetry can be identified with some Lie group $%
G$ action on the phase space $P$ and its natural extension on the whole
manifold $M$ proves to be symplectic and even more, Hamiltonian, the
Marsden-Weinstein reduction method \cite{AM,BPS} makes it possible to
construct new Hamiltonian flows on the smaller invariant reduced phase space
$\bar{M}_{\xi }:=M_{\xi }/G_{\xi }$ subject to the group invariant
constraint $p:=\xi \in \mathcal{G}^{\ast }$ for some specially chosen
element $\xi \in \mathcal{G}^{\ast }$, where $p:M\rightarrow \mathcal{G}%
^{\ast }$ is the related momentum mapping on the symplectic manifold $M$ and
$\mathcal{G}^{\ast }$ is the adjoint space to the Lie algebra $\mathcal{G}$
of the group Lie $G$.

As the corresponding Hamiltonian flows on the reduced phase space $\bar{M}%
_{\xi }$ possess often very interesting properties important for
applications in many branches of mathematics and physics, their studies were
topics of many researches during the past decades. Being interested in Lax
type flows, we observed that their modern Lie algebraic descriptions by
means of Hamiltonian group actions via the classical
Lie---Poisson---Adler---Kostant---Souriau---Berezin---Kirillov (LPAKSBK)
scheme is actually closely related to the Marsden---Weinstein reduction. In
particular, the LPAKSBK on the adjoint space $\hat{\mathcal{G}}^{\ast }$ to
the Lie algebra $\hat{\mathcal{G}}$ of a suitably chosen Lie group $G$
follows directly from an application of Marsden---Weinstein reduction to $%
M=T^{\ast }(P),$ where $P$ is chosen so that there is a naturally related
Hamiltonian group $G$-action on $M.$ Moreover, such classical integrability
theory ingredients as the $R$-structures \cite{S-T} and the related
commutation properties of the related transfer matrices are also naturally
retrieved from the Marsden-Weinstein reduction method within the scheme
specified above. These and some related aspects of this reduction technique
are topics of this investigation.

\section{Loop groups, canonically symplectic manifold and Hamiltonian action}

Consider a complex matrix Lie group $G=SL(\nu ;\mathbb{C}),\nu \in \mathbb{Z}%
_{+}$, its Lie algebra $\mathcal{G}$, and a related \cite{FT,Ne,RS-T1}
formal loop group $\tilde{G}\subset C^{\infty }(\mathbb{S}^{1};\mathrm{Hol}(%
\mathbb{C};G))$ of $G$-valued functions on the circle $\mathbb{S}^{1}$,
meromorphically depending on the complex parameter $\lambda \in \mathbb{C}$.
Its Lie algebra $\tilde{\mathcal{G}}$ can be viewed as the completion
\begin{equation}
\tilde{\mathcal{G}}=\bigcup\limits_{n\in \mathbb{Z}}\left\{ \sum_{j=-\infty
}^{n}\tilde{X}_{j}\lambda ^{j}:\ \tilde{X}_{j}\in C^{\infty }(\mathbb{S}^{1};%
\mathcal{G}),\ j\leq n\right\} .  \label{eq1.1}
\end{equation}%
Using the standard procedure \cite{FT,BPS} one can construct the centrally
extended current algebra $\hat{\mathcal{G}}:=\tilde{\mathcal{G}}\oplus
\mathbb{C}$, on which the adjoint loop group $\tilde{G}$-action is defined:
for any $g\in \tilde{G},$
\begin{equation}
g:(T,c)\rightarrow (gTg^{-1},c+(g^{-1}g_{x},T)_{-1}).  \label{eq1.2}
\end{equation}%
Here $(T,c)\in \hat{\mathcal{G}}$ \ and $(\cdot ,\cdot )_{-1}:\ \tilde{%
\mathcal{G}}\times \tilde{\mathcal{G}}\rightarrow \mathbb{C}$ is the
following nondegenerate symmetric scalar product on $\tilde{\mathcal{G}}$:
\begin{equation}
(A,B)_{-1}:=\mathrm{res}\int_{0}^{2\pi }\mathrm{tr}(A(x;\lambda )B(x;\lambda
))=(B,A)_{-1},  \label{eq1.3}
\end{equation}%
for any $A,B\in \tilde{\mathcal{G}}$. The scalar product (\ref{eq1.3}) is
ad-invariant, that is
\begin{equation}
(A,[B,C])_{-1}=([A,B],C)_{-1}  \label{eq1.4}
\end{equation}%
for any elements $A,B$ and $C\in \tilde{\mathcal{G}}.$

Define now the canonically symplectic phase space $M:=T^{\ast }(\hat{%
\mathcal{G}})\simeq (\hat{\mathcal{G}},\hat{\mathcal{G}}^{\ast })$ with the
corresponding Liouville 1-form on $M:$
\begin{equation}
\alpha ^{(1)}(T,c;l,k)=(l,dT)_{-1}+kdc,  \label{eq1.5}
\end{equation}%
whose exterior derivative gives the symplectic structure on the functional
manifold $M$:
\begin{equation}
\omega ^{(2)}(T,c;l,k):=d\alpha ^{(1)}(T,c;l,k)=(dl,\wedge dT)_{-1}+dk\wedge
dc.  \label{eq1.6}
\end{equation}%
Similarly to (\ref{eq1.2}) one can naturally extend the group $\tilde{G}$%
-action on the whole phase space $M$, having
\begin{equation}
g:(l,k)\rightarrow (glg^{-1}-kg_{x}g^{-1},k)  \label{eq1.7}
\end{equation}%
for any $(l,k)\in \hat{\mathcal{G}}^{\ast }$ and $g\in \tilde{G}$ as the
corresponding co-adjoint action of the current group $\tilde{G}$ to the
adjoint linear space $\hat{\mathcal{G}}^{\ast }.$The following lemma is
almost evident.

\begin{lemma}
\label{L1.1} The $\tilde{G} $-group action (\ref{eq1.2}) and (\ref{eq1.7})
on the symplectic phase space $M $ is symplectic and Hamiltonian.
\end{lemma}

\noindent \textbf{Proof.} \ It is easy to check that the canonical Liouville
1-form (\ref{eq1.5}) on the manifold $M$ is $\tilde{G}$-invariant:
\begin{eqnarray}
&&  \notag \\
&&g^{\ast }\alpha
^{(1)}(T,c;l,k)=(glg^{-1}-kg_{x}g^{-1},gdTg^{-1})_{-1}+k(dc+(g^{-1}g_{x},dT)_{-1})=
\notag \\
&=&(glg^{-1},gdTg^{-1})-k(g_{x}g^{-1},gdTg^{-1})_{-1}+kdc+k(g^{-1}g_{x},dT)_{-1}=
\label{eq1.8} \\
&=&(l,g^{-1}gdTg^{-1}g)_{-1}-k(g^{-1}g_{x}g^{-1}g,dT)_{-1}+kdc+k(g^{-1}g_{x},dT)_{-1}=
\notag \\
&=&(l,dT)_{-1}+kdc=\alpha ^{(1)}(T,c;l,k).  \notag \\
&&  \notag
\end{eqnarray}%
From (\ref{eq1.8}), owing to the expression (\ref{eq1.6}), one obtains the
symplectic form invariance
\begin{equation}
g^{\ast }\omega ^{(2)}(T,c;l,k)=\omega ^{(2)}(T,c;l,k)  \label{eq1.9}
\end{equation}%
for any element $(T,c;l,k)\in M.$

To define the Hamiltonian $\tilde{G}$-action on the symplectic manifold $M$
we take the group flow $g(t):=\exp (tX)$ for $t\in \mathbb{R}$, $X\in \tilde{%
\mathcal{G}}$, and find the driven generated vector field $%
K_{X}:M\rightarrow T(M)$ on the phase space $M$:
\begin{eqnarray}
&&K_{X}(T,c;l,k):=  \label{eq1.10} \\
&=&\frac{d}{dt}%
(g(t)Tg(t)^{-1},c+(g(t)^{-1}g_{x}(t),T)_{-1};g(t)lg(t)^{-1}-kg_{x}(t)g(t)^{-1},k)%
\bigg|_{t=0}=  \notag \\
&=&([X,T],(X_{x},T)_{-1};[X,l]-kX_{x},0),  \notag \\
&&  \notag
\end{eqnarray}%
by a Hamiltonian function $H_{X}:M\rightarrow \mathbb{C}$ owing to the
canonical relationship $-dH_{X}=i_{K_{X}}\omega ^{(2)}:$
\begin{eqnarray}
&&-dH_{X}=-(\partial H/\partial l,dl)_{-1}-(\partial H_{X}/\partial
T,dT)_{-1}+  \label{eq1.11} \\
&&+\partial H_{X}/\partial k\ dk+\partial H_{X}/\partial c\ dc=  \notag \\
&=&([X,l]-kX_{x},dT)_{-1}-(dl,[X,T])_{-1}-(X_{x},T)_{-1}dk.  \notag
\end{eqnarray}%
As a consequence of (\ref{eq1.11}) one obtains
\begin{eqnarray}
&&\partial H_{X}/\partial l=[X,T],\ \ \ \ \ \partial H_{X}/\partial
T=kX_{x}-[X,l],  \label{eq1.12} \\
&&\partial H_{X}/\partial k=(X_{x},T)_{-1},\ \ \ \ \partial H_{X}/\partial
c=0  \notag
\end{eqnarray}%
for any point $(T,k;l,c)\in M.$ From (\ref{eq1.12}) it follows that
\begin{equation}
H_{X}=([T,l]-kT_{x},X)_{-1}:=(p(T,c;l,k),X)_{-1},  \label{eq1.13}
\end{equation}%
is linear with respect to the generator element $X\in \tilde{\mathcal{G}}.$
This means that the loop group $\tilde{\mathcal{G}}^{\ast }$ action on the
symplectic manifold $M$ is Hamiltonian by definition \cite{AM,PM}. $%
\triangleright $

The corresponding mapping $p:M\rightarrow \tilde{\mathcal{G}}^{\ast },$
where
\begin{equation}
p(T,c;l,k)=[T,l]-kT_{x},  \label{eq1.14}
\end{equation}%
is called the momentum mapping \cite{AM,BPS,PM} which can be constrained to
be fixed for further applications to the phase space $M$ in the
Marsden-Weinstein reduction procedure \cite{AM}.

Let us describe in detail the related symplectic structure on the $\xi $%
-level submanifold
\begin{equation}
M_{\xi }:=\left\{ (T,c;l,k)\in M:\ [T,l]-kT_{x}=\xi \in \tilde{\mathcal{G}}%
^{\ast }\right\}  \label{eq1.15}
\end{equation}%
for a fixed element $\xi \in \tilde{\mathcal{G}}^{\ast }.$ As a more natural
case we take that $\xi =0\in \tilde{\mathcal{G}}^{\ast }.$ The corresponding
isotropy group $\tilde{G}_{\xi }=\tilde{G}$, as $\mathrm{Ad}_{g}^{\ast }\xi %
\big|_{\xi =0}=0$ holds for any element $g\in \tilde{G}.$ To proceed further
we need some additional properties of the submanifold $M_{\xi }\subset M,$
which we will describe in the next section

\section{Marsden-Weinstein reduction, commuting vector fields and Poisson
bracket}

In this section we will be interested in describing the submanifold $M_{\xi
}\subset {M}$ parameterized by the points of the reduced phase space $\bar{M}%
_{\xi }:=M_{\xi }/G_{\xi }$. It is known \cite{AM,Ar}, that this
parametrization uniquely determines the points $(\bar{T},\bar{c};\bar{l},%
\bar{k})\in M_{\xi }\subset M,$which are invariant with respect to the
appropriate loop group $\tilde{G}$ action (\ref{eq1.2}) and (\ref{eq1.7}).
The last property makes it possible \cite{AM,Ar,BPS,PSP} to define on the
phase space $\bar{M}_{\xi }$ the reduced nondegenerate symplectic structure
on the phase space $\bar{M}_{\xi }$ by means of the appropriate symplectic
structure on the submanifold $M_{\xi }$. Let us consider the point $(\bar{T},%
\bar{c};\bar{l},\bar{k})\in M_{\xi },$ where the elements $\bar{T}\in \tilde{%
\mathcal{G}}$ $,\bar{k}\in \mathbb{C},$ according to the definition (\ref%
{eq1.15}), satisfy the differential expressions:
\begin{equation}
\lbrack \bar{T},\bar{l}]-\bar{k}\bar{T}_{x}=0,\ \ \ \bar{k}_{x}=0,
\label{eq2.4a}
\end{equation}%
for all $\ x\in \mathbb{S}^{1}.$ Consider now a Hamiltonian vector field $-%
\bar{k}d/d\tau ,\tau \in \mathbb{C},$ on the submanifold $M_{\xi },$
generated by the element $X=\bar{l}\in \tilde{\mathcal{G}}^{\ast }$ owing to
the expressions
\begin{equation}
-\bar{k}\bar{T}_{\tau }=[\bar{l},\bar{T}]=-[\bar{T},\bar{l}]=-\bar{k}\bar{T}%
_{x},\ \ \ \ -\bar{k}\bar{l}_{\tau }=\bar{k}\bar{l}_{x}.  \label{eq2.4}
\end{equation}%
From (\ref{eq2.4}) it follows that the equality $\frac{d}{d\tau }=\frac{d}{dx%
}$ holds on the reduced phase space $\bar{M}_{\xi }.$ Let us compute
additionally the evolution of the element $\bar{c}\in \mathbb{C}$ with
respect to this vector field $d/d\tau $ on $\bar{M}_{\xi }:$
\begin{equation}
-\bar{k}\bar{c}_{\tau }=(\bar{l}_{x},\bar{T})_{-1}=-(\bar{l},\bar{T}%
_{x})_{-1}=-(\bar{l},\bar{k}^{-1}[\bar{T},\bar{l}])_{-1}=\bar{k}^{-1}([\bar{l%
},\bar{l}],\bar{T})_{-1}=0,  \label{eq2.5}
\end{equation}%
coinciding with the \textit{a priori }assumed condition $d\bar{c}/dx=0$ for
any $x\in \mathbb{S}^{1}.$

Define similarly a vector field $d/dt,$ $t\in \mathbb{C},$ on the reduced
phase space $\bar{M}_{\xi },$ generated by the Lie algebra element $q(\bar{l}%
)\in \tilde{\mathcal{G}},$ depending on the basis element $\bar{l}\in \tilde{%
\mathcal{G}}^{\ast }$ such \ that
\begin{equation}
\bar{T}_{t}=[q(\bar{l}),\bar{T}],\ \ \ \bar{l}_{t}=[q(\bar{l}),\bar{l}]-\bar{%
k}\bar{l}_{x},\ \ \ \bar{c}_{t}=(q_{x}(\bar{l}),\bar{T})_{-1},\ \ \ \ \bar{k}%
_{t}=0.  \label{eq2.6}
\end{equation}%
The latter, in particular, means that the flows $d/dt$ and $d/dx$ on the
reduced phase space $\bar{M}_{\xi }$ possess the countable set $\gamma _{n}(%
\bar{l}):=\mathrm{tr}\bar{T}^{n}(\bar{l}),$ $n\in \mathbb{Z},$ of
conservation lows, where by definition, the element $\bar{T}(\bar{l})\in
\tilde{\mathcal{G}}$ satisfies for a given element $\bar{l}\in \tilde{%
\mathcal{G}}^{\ast }$ the determining equation
\begin{equation}
-\bar{k}\bar{T}_{x}(\bar{l})=[\bar{l},\bar{T}(\bar{l})]  \label{eq2.6a}
\end{equation}%
for all $x\in \mathbb{S}^{1}.$ From the equations (\ref{eq2.6a}) one easily
finds that upon the reduced phase space $\bar{M}_{\xi }$
\begin{eqnarray}
&&\bar{c}_{t}=(q(\bar{l})_{x},\bar{T})_{-1}=\bar{k}^{-1}([q(\bar{l}),\bar{l}%
]-\bar{l}_{t},\bar{T})_{-1}=  \notag \\
&=&\bar{k}^{-1}([q(\bar{l}),\bar{l}],\bar{T})_{-1}-\bar{k}^{-1}(\bar{l}_{t},%
\bar{T})=\bar{k}^{-1}([\bar{T},q(\bar{l})],\bar{l})_{-1}-  \notag \\
&&-\bar{k}^{-1}(\bar{l}_{t},\bar{T})_{-1}=-\bar{k}^{-1}(\bar{l},\bar{T}%
_{t})_{-1}-\bar{k}^{-1}(\bar{l}_{t},\bar{T})_{-1}=  \notag \\
&&-\bar{k}^{-1}\frac{d}{dt}(\bar{l},\bar{T})_{-1}.  \label{eq2.7}
\end{eqnarray}%
Thus, from the $t$-evolution (\ref{eq2.7}) of the parameter $\bar{c}\in
\mathbb{C}$ one finds that the constraint
\begin{equation}
\bar{c}=-\bar{k}^{-1}(\bar{l},\bar{T})_{-1}  \label{eq2.8}
\end{equation}%
holds on the reduced phase space $\bar{M}_{\xi }$ subject to the vector
field $d/dt$ generated by the element $q(\bar{l})\in \tilde{\mathcal{G}}.$
Moreover, as it is easy to observe, these two vector fields $d/d\tau $ and $%
d/dt$ on the reduced phase space $\bar{M}_{\xi }$ commute:
\begin{equation}
\lbrack d/dt,d/d\tau ]=0.  \label{eq2.9}
\end{equation}%
The latter is very promising, since the condition (\ref{eq2.9}) results in
some differential relationships on the components of the reduced matrix $%
\bar{l}\in \tilde{\mathcal{G}}^{\ast },$ \ for which the related linear
evolution equation
\begin{equation}
\bar{F}_{x}=\bar{l}\bar{F},  \label{eq2.9a}
\end{equation}%
augmented with the compatible differential equation
\begin{equation}
\bar{F}_{t}=q(\bar{l})\bar{F}  \label{eq2.10}
\end{equation}%
for the matrix $F\in \tilde{G}$ \ are compatible. These equations (\ref%
{eq2.9a}) and (\ref{eq2.10}) realize the well known \cite%
{FT,No,Ne,RS-T1,PM,BPS} generalized Lax type spectral problem, allowing to
integrate the mentioned above differential relationships by means of either
the inverse scattering or the spectral transform methods \cite{FT,No,Ne,CD}
and algebraic geometry methods \cite{No,Ne}, or their modern generalizations
\cite{RS-T1}.

To make this aim more constructive, it is necessary to describe the
evolution of the vector field $d/dt$ on the reduced phase space $\bar{M}%
_{\xi }$ in more detail subject to its dependence on the phase space element
$\bar{l}\in \tilde{\mathcal{G}}^{\ast }.$ Taking into account that the
vector fields $d/dt$ and $d/dx$ satisfy the commutation condition (\ref%
{eq2.9}) on the reduced manifold $M_{\xi }$, we will apply the
Marsden-Weinstein reduction theory to our symplectic manifold $M$ with the
fixed value of the moment mapping $\xi =0$ for computing the basic Poisson
bracket
\begin{equation}
\left\{ (\bar{T},X)_{-1},(\bar{T},Y)_{-1}\right\} _{\xi }  \label{eq2.10a}
\end{equation}%
of the functions $(\bar{T},X)$ and $(\bar{T},Y)$ on the reduced phase space $%
\bar{M}_{\xi }$ for arbitrary $X,Y\in \tilde{\mathcal{G}}^{\ast }.$ \ It can
be shown \cite{ABT,PSP,BPS} that this Poisson bracket on $\bar{M}_{\xi }$ in
general is
\begin{equation}
\left\{ (\bar{T},X)_{-1},(\bar{T},Y)_{-1}\right\} _{\xi }=\left\{ (\bar{T}%
,X)_{-1},(\bar{T},Y)_{-1}\right\} \big|_{\bar{M}_{\xi }}-(\xi
,[V_{X},V_{Y}])_{-1}\big|_{\bar{M}_{\xi }},  \label{eq2.11}
\end{equation}%
where, by definition, the mappings $V_{X},V_{Y}:\bar{M}_{\xi }\rightarrow
\mathcal{\tilde{G}}$ denote the solutions to the following relationship:
\begin{equation}
(\xi ,[Z,V_{X}])_{-1}=K_{Z}(T,X)_{-1},(\xi ,[Z,V_{Y}])_{-1}=K_{Z}(T,Y)_{-1},
\label{eq2.12}
\end{equation}%
which holds for all $Z\in \tilde{\mathcal{G}.}\ $ The functions $(\bar{T}%
,X)_{-1},(\bar{T},Y)_{-1}\in \mathcal{D}(\bar{M}_{\xi })$ should be extended
to those on the whole phase space $M$ in such a way that their restrictions
on the submanifold $M_{\xi }\subset M$ \ are $\tilde{G}$-invariant.

To apply the Marsden-Weinstein reduction we will take into account that, by
definition, there exists a group element $g(l)\in \tilde{G}$ such that for
arbitrarily chosen $l\in \tilde{G}$ the expression
\begin{equation}
l=g(l)\bar{l}(l)g(l)^{-1}-\bar{k}{g}_{x}(l)g(l)^{-1}  \label{eq2.13}
\end{equation}%
holds and satisfies the normalization condition $g(\bar{l})=\mathrm{Id}\in
\tilde{G}$. By considering the function
\begin{equation}
f_{X}:=(T,g(l)Xg(l)^{-1})_{-1},  \label{eq2.14}
\end{equation}%
one can observe that $f_{X}|_{\bar{M}_{\xi }}=(\bar{T},X)_{-1}$ and, by
construction, it is $\tilde{G}$-invariant. The latter means that $f_{X}\in
\mathcal{D}(M_{\xi })$ for any $l\in \tilde{\mathcal{G}}^{\ast }.$ \ In
fact, for any $a\in \tilde{G}_{\xi }=\tilde{G}$
\begin{eqnarray}
&&a\circ f_{X}:=(a\cdot T,g(a\circ l)Xg(a\circ l)^{-1})_{-1}=  \label{eq2.15}
\\
&=&(aTa^{-1},ag(l)Xg(l)^{-1}\cdot a^{-1})=(T,g(l)Xg(l)^{-1})_{-1}=f_{X},
\notag
\end{eqnarray}%
where we made use of the property $g(a\circ l)=a\ g(l),$ $l\in \tilde{%
\mathcal{G}}^{\ast }.$ The latter holds owing to the definitions (\ref%
{eq2.13}) and (\ref{eq1.7}):
\begin{eqnarray}
&&  \notag \\
&&a\circ l=ala^{-1}-\bar{k}a_{x}a^{-1}=a(g(l)\bar{l}g(l)^{-1}-\bar{k}%
g_{x}(l)g(l)^{-1})a^{-1}-\bar{k}a_{x}a^{-1}=  \notag \\
&=&ag(l)\bar{l}(ag(l))^{-1}-\bar{k}ag_{x}(l)g(l)^{-1}a^{-1}-\bar{k}%
a_{x}a^{-1}=  \notag \\
&=&ag(l)\bar{l}(ag(l))^{-1}-\bar{k}(ag(l))_{x}(ag(l))^{-1}=  \label{2.16} \\
&=&g(a\circ l)\bar{l}g(a\circ l)^{-1}-\bar{k}g_{x}(a\circ l)g(a\circ l)^{-1},
\notag \\
&&  \notag
\end{eqnarray}%
giving rise to relationship $g(a\circ l)=a$ $g(l)$ for any $a\in \tilde{G}%
_{\xi }$ and $l\in \tilde{\mathcal{G}}^{\ast }.$

Returning to the Poisson bracket (\ref{eq2.11}), we can replace the
functions $(\bar{T},X)_{-1}$ and $(\bar{T},Y)_{-1}\in \mathcal{D}(\bar{M}%
_{\xi })$ with their $\tilde{G}_{\xi }$-invariant extensions $f_{X}\in
\mathcal{D}(M_{\xi }).$ Before calculating the corresponding Poisson bracket
\begin{equation}
\left\{ \bar{f}_{X},\bar{f}_{Y}\right\} _{\xi }=\left\{ \bar{f}_{X},\bar{f}%
_{Y}\right\} |_{\bar{M}_{\xi }}-(\xi ,[V_{X},V_{Y}])_{-1}=\left\{ {f}_{X},{f}%
_{Y}\right\} |_{\bar{M}_{\xi }}-K_{V_{X}}f_{Y}|_{\bar{M}_{\xi }},
\label{eq2.17}
\end{equation}%
where $K_{V_{X}}:M\rightarrow T(M)$ is the vector field generated on $M$ by
the element $V_{X}\in \tilde{\mathcal{G}},$ we need to calculate the action $%
K_{Z}f_{Y}$ for any element $Z\in \tilde{\mathcal{G}}.$ Similarly to the
calculations from \cite{ABT}, one finds that on the submanifold $M_{\xi }$
\begin{eqnarray}
&&K_{Z}f_{Y}=\frac{d}{d\varepsilon }\left( \exp (\varepsilon Z)T\exp
(-\varepsilon Z),g(\exp (\varepsilon Z)\circ l)Yg(\exp (\varepsilon Z)\circ
l)^{-1}\right) _{-1}|_{\varepsilon =0}=  \label{eq2.18} \\
&=&(T,g(l)[g(l)^{-1}g^{\prime }(l)([Z,l]-\bar{k}%
Z_{x})-g(l)^{-1}Zg(l),Y]g(l)^{-1})_{-1}.  \notag
\end{eqnarray}%
Thus, on the reduced phase space $\bar{M_{\xi }}$ the general expression (%
\ref{eq2.18}) implies
\begin{equation}
K_{V_{X}}f_{Y}|_{\bar{M}_{\xi }}=(\bar{T},[g^{\prime }(\bar{l})\cdot ([V_{x},%
\bar{l}]-\bar{k}\frac{d}{dx}V_{x})-V_{X},Y])_{-1}.  \label{eq2.19}
\end{equation}%
Thus, the Poisson bracket (\ref{eq2.17}), owing to the relationships $%
\left\{ f_{X},f_{Y}\right\} =-\omega ^{(2)}(K_{V_{X}},K_{V_{Y}})$ and (\ref%
{eq2.19}), becomes

\begin{eqnarray}
&&\left\{ (\bar{T},X),(\bar{T},Y)\right\} _{\xi }=  \label{eq2.20} \\
&&\left( \bar{T},[g^{\prime }(\bar{l})(Y),X]+[Y,g^{\prime }(\bar{l}%
)(X)]\right) _{-1}-\left( \bar{T},[g^{\prime }(\bar{l})([V_{X},\bar{l}]-\bar{%
k}\frac{d}{dx}V_{X})-V_{X},Y]\right) _{-1}=  \notag \\
&=&(\bar{T},[g^{\prime }(\bar{l})(Y),X]+[Y,g^{\prime }(\bar{l})(X)])_{-1},
\notag \\
&&  \notag
\end{eqnarray}%
where we take into account that owing to (\ref{eq2.12}) and (\ref{eq2.19}),
the expression
\begin{equation*}
\left( \bar{T},[g^{\prime }(\bar{l})([V_{X},\bar{l}]-\bar{k}\frac{d}{dx}%
V_{X})-V_{X},Y]\right) _{-1}=K_{V_{X}}f_{Y}=(\xi ,[K_{V_{X}},V_{Y}])_{-1}%
\big{|}_{\xi =0}=0.
\end{equation*}%
Now one can rewrite the Poisson bracket (\ref{eq2.20}) as
\begin{equation}
\left\{ (\bar{T},X),(\bar{T},Y)\right\} _{\xi }=(\bar{T},[X,Y]_{D})_{-1},
\label{eq2.21}
\end{equation}%
where, by definition, we have introduced the classical $D$-matrix structure
in the Lie algebra $\tilde{\mathcal{G}}^{\ast }$:
\begin{equation}
\lbrack X,Y]_{D}:=[D(X),Y]+[X,D(Y)],  \label{eq2.22}
\end{equation}%
where $X,Y\in \tilde{\mathcal{G}}^{\ast }$ and the linear homomorphism $D:%
\tilde{\mathcal{G}}^{\ast }\rightarrow \tilde{\mathcal{G}}^{\ast }$ is
defined as
\begin{equation}
D(X):=-g^{\prime }(\bar{l})(X).  \label{eq2.23}
\end{equation}%
The mapping (\ref{eq2.23}) should satisfy \cite{BV} the well known condition
\begin{equation}
(\bar{T},[X,[D(Y),D(Z)]-D[Y,Z]_{D}])_{-1}+(\bar{T},[X,\{(\bar{T},Y),(\bar{T}%
,Z)\}])+\mathrm{cycles}=0  \label{eq2.24}
\end{equation}%
for any $X,Y\in \tilde{\mathcal{G}}^{\ast }$ and $Z\in \tilde{\mathcal{G}}.$

Now it is useful to recall that the mapping $g:\tilde{\mathcal{G}}^{\ast
}\rightarrow \tilde{G}$ satisfies the relationship (\ref{eq2.13}), which
implies \cite{ArMe} the following differential expression
\begin{equation}
\lbrack g^{\prime }(\bar{l})(X),\bar{l}]-\bar{k}\frac{d}{dx}g^{\prime }(\bar{%
l})(X)=X  \label{eq2.25}
\end{equation}%
for any $X\in \tilde{\mathcal{G}}^{\ast },$ \ where $g^{\prime }(\bar{l}):%
\tilde{\mathcal{G}}^{\ast }\rightarrow \tilde{\mathcal{G}}^{\ast }$ is the
derivative mapping, depending from the chosen reduction ${\mathcal{G}}^{\ast
}\ni {l}\rightarrow \bar{l}\ \in {\mathcal{G}}^{\ast }.$

The mapping (\ref{eq2.23}) satisfies an additional relationship, which can
be obtained from the group $\tilde{G}$-action on the element $\bar{T}(\bar{l}%
)\in \tilde{\mathcal{G}}:$
\begin{equation}
T(l)=g(l)\bar{T}(\bar{l})g(l)^{-1},  \label{eq2.26}
\end{equation}%
following naturally from (\ref{eq2.13}). Differentiation of (\ref{eq2.26})
with respect to $l\in \tilde{\mathcal{G}}^{\ast }$ at the point $l=\bar{l},$
gives rise to the expression
\begin{equation}
T^{\prime }(\bar{l})(X)=[g^{\prime }(\bar{l})(X),\bar{T}(\bar{l})]
\label{eq2.27}
\end{equation}%
for an arbitrary $X\in \tilde{\mathcal{G}}^{\ast }.$ Moreover, since the
matrix (\ref{eq2.26}) satisfies the relationship (\ref{eq2.6a}), its
differentiation with respect to $\bar{l}\in \tilde{\mathcal{G}}^{\ast }$
entails the differential expression:
\begin{equation}
\bar{k}\frac{d}{dx}T^{\prime }(\bar{l})(Y)+[\bar{l},T^{\prime }(\bar{l}%
)(Y)]=[\bar{T}(\bar{l}),Y],  \label{eq2.28}
\end{equation}%
which holds for any $Y\in \tilde{\mathcal{G}}^{\ast }$. The above results
can be formulated as the following proposition.

\begin{proposition}
The Poisson bracket (\ref{eq2.10a}) on the reduced phase space $\bar{M}_{\xi
}$ represented as a $D$-structure (\ref{eq2.21}) on the linear space $\tilde{%
\mathcal{G}}^{\ast }$, naturally generated by the gauge transformation (\ref%
{eq2.13}), which reduces the arbitrary element $l\in \tilde{\mathcal{G}}%
^{\ast }$ to the element $\bar{l}\in \tilde{\mathcal{G}}^{\ast },$ is
uniquely defined on $\bar{M}_{\xi }.$
\end{proposition}

As a consequence of representation (\ref{eq2.21}) we find that there exists
an infinite hierarchy of mutually commuting other functionals with respect
to the Poisson bracket on the phase space $\bar{M}_{\xi }$. The latter
follows from the tensor form of the Poisson bracket (\ref{eq2.10a}) in the
space $\tilde{\mathcal{G}}\otimes \tilde{\mathcal{G}}:$
\begin{equation}
\left\{ \bar{T}(\bar{l})(\lambda )\overset{\otimes }{,}\bar{T}(\bar{l})(\mu
)\right\} _{\xi }=[D(\lambda ,\mu ),\bar{T}(\bar{l})(\lambda )\otimes
\mathbb{I}+\mathbb{I}\otimes \bar{T}(\bar{l})(\mu )]  \label{eq2.29a}
\end{equation}%
which holds for arbitrary $\lambda ,\mu \in \mathbb{C}$ and where $D(\lambda
,\mu ):\tilde{\mathcal{G}}^{\ast }\rightarrow \tilde{\mathcal{G}}^{\ast }$
denotes the tensor form of the $D$-structure $D:\tilde{\mathcal{G}}^{\ast
}\rightarrow \tilde{\mathcal{G}}^{\ast }.$ The trace operation in (\ref%
{eq2.29a}) causes the Poisson bracket vanish on the phase space $\bar{M}%
_{\xi }$ for the functionals $\mathrm{tr}\bar{T}(\bar{l})(\lambda )$ and $%
\mathrm{tr}\bar{T}(\bar{l})(\mu )$ for arbitrary $\lambda ,\mu \in \mathbb{C}%
.$

\section{Monodromy matrix, associated $R$-structure and Lie-Poisson bracket}

Next we analyze possible forms of the $D$-mapping (\ref{eq2.23}) as a
function on the reduced phase space $\bar{M}_{\xi }.$ Since the parameter $%
\bar{k}\in \mathbb{C}$ is constant, its value for convenience is set at $%
\bar{k}=-1.$ Thus, taking into account the definition (\ref{eq2.23}), the
determining $D$-structure equation (\ref{eq2.25}) takes the form:
\begin{equation}
\lbrack D(\bar{l})(Y),\bar{l}]+\frac{d}{dx}D(\bar{l})(Y)+Y=0  \label{eq2.28c}
\end{equation}%
for any element $Y\in \tilde{\mathcal{G}}^{\ast }.$

Let us consider the linear matrix equation
\begin{equation}
\bar{F}_{x}(x,s;\lambda )=\bar{l}(x;\lambda )\bar{F}(x,s;\lambda ),
\label{eq2.1}
\end{equation}%
where $\bar{l}(x;\lambda )\in \tilde{\mathcal{G}}^{\ast },\bar{F}\in \tilde{G%
},$ with Cauchy data at a point $x=s\in \mathbb{S}^{1}:$
\begin{equation}
\bar{F}(x,s;\lambda )\big|_{x=s}=\mathbb{I}.  \label{eq2.2}
\end{equation}%
The corresponding normalized monodromy matrix
\begin{equation}
{\bar{T}(x;\lambda ):=\bar{F}(x+2\pi ,x;\lambda )-\nu ^{-1}\mathbb{I}\mathrm{%
tr}\bar{F}(x+2\pi ,x;\lambda ),}  \label{eq2.2a}
\end{equation}%
for $x\in \mathbb{S}^{1}$ and arbitrary $\lambda \in \mathbb{C}$ satisfies
the differential expression
\begin{equation}
\bar{T}_{x}-[\bar{T},\bar{l}]=0,  \label{eq2.3}
\end{equation}%
exactly coinciding with (\ref{eq2.6a}). Thus, if by means of the co-adjoint
transformation (\ref{eq1.7}) this chosen matrix ${l}\in \tilde{\mathcal{G}}%
^{\ast }$ will be transformed into the matrix $\bar{l}\in \tilde{\mathcal{G}}%
^{\ast },$ then the corresponding monodromy matrix of the equation (\ref%
{eq2.9a}) will transform into the monodromy matrix of the equation (\ref%
{eq2.1}), which satisfies the expression (\ref{eq2.3}).

Taking into account the differential relationships (\ref{eq2.1}), (\ref%
{eq2.2}) and (\ref{eq2.3}), one can recalculate the Poisson bracket (\ref%
{eq2.21}) by means of the identification
\begin{equation}
\bar{T}(\bar{l})(z;\lambda )=\bar{T}(z;\lambda )  \label{eq2.29}
\end{equation}%
for arbitrary $z\in \mathbb{S}^{1}$ and $\lambda \in \mathbb{C}^{1}.$ It
yields the following tensor expression for the reduced phase space $\ \bar{M}%
_{\xi }:$
\begin{eqnarray}
&&  \label{eq2.30} \\
&&\left\{ \bar{T}(\bar{l})(z;\lambda )\overset{\otimes }{,}\bar{T}(\bar{l}%
)(z;\mu )\right\} _{\xi }=  \notag \\
&=&\int\limits_{z}^{z+2\pi }dx\int\limits_{z}^{z+2\pi }dy\left\{ F(z+2\pi
,x;\lambda )\bar{l}(x;\lambda )F(x,z;\lambda )\overset{\otimes }{,}F(z+2\pi
,y;\mu )\bar{l}(y;\mu )F(y,z;\mu )\right\} _{\xi }=  \notag \\
&=&\int\limits_{z}^{z+2\pi }dx\int\limits_{z}^{z+2\pi }dy\left\{ (F(z+2\pi
,x;\lambda )\otimes \mathbb{I})(\bar{l}(x;\lambda )\otimes \mathbb{I}%
)(F(x,z;\lambda )\otimes \mathbb{I},\right.  \notag \\
&&\left. \mathbb{I}\otimes F(z+2\pi ,y;\mu )(\mathbb{I}\otimes \bar{l}(y;\mu
))(\mathbb{I}\otimes F(y,z;\mu ))\right\} _{\xi }=  \notag \\
&=&\int\limits_{z}^{z+2\pi }dx\int\limits_{z}^{z+2\pi }dyF(z+2\pi ,x;\lambda
)\otimes F(z+2\pi ,y;\mu )\left\{ \bar{l}(x;\lambda )\overset{\otimes }{,}%
\bar{l}(y;\mu )\right\} _{\xi }F(x,z;\lambda )\otimes F(y,z;\mu )  \notag \\
&&\int\limits_{z}^{z+2\pi }dx\int\limits_{z}^{z+2\pi }dyF(z+2\pi ,x;\lambda
)\otimes F(z+2\pi ,y;\mu )\bar{\omega}(\lambda ,\mu ;x,y)F(x,z;\lambda
)\otimes F(y,z;\mu ),  \notag \\
&&
\end{eqnarray}%
where $z\in \mathbb{S}^{1}$, $\lambda ,\mu \in \mathbb{C}$ and, by
definition,
\begin{equation}
\left\{ \bar{l}(x;\lambda )\overset{\otimes }{,}\bar{l}(y;\mu )\right\}
_{\xi }:=\bar{\omega}(\lambda ,\mu ;x,y)=\sum\limits_{i,k=0}^{N}\bar{\omega}%
_{ik}(\lambda ,\mu ;x,y)\partial _{x}^{i}\partial _{y}^{k}\delta (x-y).
\label{eq2.31}
\end{equation}%
Here the local functional matrices $\bar{\omega}_{ik}(\lambda ,\mu ;x,y)\in
\tilde{\mathcal{G}}^{\ast }\otimes \tilde{\mathcal{G}}^{\ast }$ satisfy the
antisymmetry property:
\begin{equation}
P\bar{\omega}_{ik}(\lambda ,\mu ;x,y)P=-\bar{\omega}_{ki}(\mu ,\lambda ;x,y)
\label{eq2.32_1}
\end{equation}%
for all $i,k=\overline{1,N},$ $x,y\in \mathbb{S}^{1},$ $\lambda ,\mu \in
\mathbb{C}$ and the permutation operator $P:\tilde{\mathcal{G}}^{\ast
}\otimes \tilde{\mathcal{G}}^{\ast },$ acts as $PA\otimes BP:=B\otimes A$
for any $A,B\in \tilde{\mathcal{G}}^{\ast }.$ Just as in the calculation
from \cite{FT,Ts,Sk} one obtains from (\ref{eq2.31}) that
\begin{equation}
\left\{ \bar{T}(z;\lambda )\overset{\otimes }{,}\bar{T}(z;\mu )\right\}
_{\xi }=\int\limits_{z}^{z+2\pi }dx\bar{F}(z+2\pi ,x;\lambda )\otimes \bar{F}%
(z+2\pi ,x;\mu )\bar{\Omega}(\lambda ,\mu ;x)\bar{F}(x,z;\lambda )\otimes
\bar{F}(x,z;\mu ),  \label{eq2.31_1}
\end{equation}%
where the matrix $\bar{\Omega}(\lambda ,\mu ;x)\in \tilde{\mathcal{G}}^{\ast
}\otimes \tilde{\mathcal{G}}^{\ast }$ for all $\lambda ,\mu \in \mathbb{C}%
,x\in \mathbb{S}^{1},$ depends only on ${\bar{l}\in \mathcal{G}}^{\ast }.$

The expression (\ref{eq2.31_1}) allows the very compact representation
\begin{equation}
\left\{ \bar{T}(z;\lambda )\overset{\otimes }{,}\bar{T}(z;\mu )\right\}
_{\xi }=\mathcal{R}(\lambda ,\mu ;z)\bar{T}(z;\lambda )\otimes \bar{T}(z;\mu
)-\bar{T}(z;\lambda )\otimes \bar{T}(z;\mu )\mathcal{R}(\lambda ,\mu ;z),
\label{eq2.34}
\end{equation}%
if the tensor $\mathcal{R}$-matrix $\mathcal{R}\in \tilde{\mathcal{G}}%
\otimes \tilde{\mathcal{G}}^{\ast }$ satisfies for $x\in \mathbb{S}^{1}$ and
$\lambda ,\mu \in \mathbb{C}$ the differential relationship
\begin{equation}
\frac{d}{dx}\mathcal{R}(\lambda ,\mu ;x)+[\mathcal{R}(\lambda ,\mu
;x),l(x;\lambda )\otimes \mathbb{I}+\mathbb{I}\otimes l(x;\mu )]=\Omega
(\lambda ,\mu ;x).  \label{eq2.35}
\end{equation}%
If we define the mapping ${R}:\tilde{\mathcal{G}}\rightarrow \tilde{\mathcal{%
G}}$ as
\begin{equation}
{R}(Y):=\underset{\mu =0}{\mathrm{res}}\int\limits_{0}^{2\pi }dy\mathcal{R}%
(\lambda ,\mu ;y)\delta (x-y)Y(y;\mu )  \label{eq2.36}
\end{equation}%
for any $Y\in \tilde{\mathcal{G}}^{\ast }$, then the relationship (\ref%
{eq2.35}) can be easily presented in the following operator form:
\begin{equation}
-(X,\frac{d{R}}{dx}(Y))_{-1}+(\bar{l},[X,Y]_{{R}})=(X,{R}(Y))_{-1},
\label{eq2.37}
\end{equation}%
which holds for any $X,Y\in \tilde{\mathcal{G}}$, where we denoted
\begin{equation}
\lbrack X,Y]_{R}:=[-{R}^{\ast }(x),Y]+[X,R(Y)].  \label{eq2.37a}
\end{equation}%
The result (\ref{eq2.37}) can be used for rewriting the Poisson bracket (\ref%
{eq2.34}) as
\begin{eqnarray}
&&  \label{eq2.38} \\
&&\left\{ (X,\bar{T}(\bar{l}))_{-1},(Y,\bar{T}(\bar{l}))_{-1}\right\} _{\xi
}=  \notag \\
&=&\left( \bar{l},[\bar{F}X\bar{F}_{2\pi },\bar{F}Y\bar{F}_{2\pi
}]_{R}\right) _{-1}-\left( \bar{F}X\bar{F}_{2\pi },\frac{R}{dx}(\bar{F}Y\bar{%
F}_{2\pi })\right) _{-1}=  \notag \\
&=&\left( \bar{l},[\nabla (X,\bar{T})(\bar{l}),\nabla (Y,\bar{T})(\bar{l}%
)]_{R}\right) _{-1}-\left( \nabla (X,\bar{T})(\bar{l}),\frac{d}{dx}R(\nabla
(Y,\bar{T})(\bar{l}))\right) _{-1}-  \notag \\
&&-\left( R^{\ast }(\nabla (X,\bar{T})(\bar{l})),\frac{d}{dx}(\nabla (Y,\bar{%
T})(\bar{l}))\right) _{-1},  \notag \\
&&  \notag
\end{eqnarray}%
where $\bar{F}:=\bar{F}(\bar{l})(x,y;\lambda ),\ \bar{F}_{2\pi }:=\bar{F}(%
\bar{l})(y+2\pi ,x;\lambda )\in \tilde{G}$ $,$ $x,y\in \mathbb{S}%
^{1},\lambda \in \mathbb{C},$ and we defined the gradients $\nabla (X,\bar{T}%
)(\bar{l})$ and $\nabla (Y,\bar{T})(\bar{l})\in \tilde{\mathcal{G}}$ \ in
the standard way as
\begin{equation}
(\nabla f(\bar{l}),Z)_{-1}:=\frac{d}{d\varepsilon }f(\bar{l}+\varepsilon Z)%
\bigg|_{\varepsilon =0}  \label{eq2.39}
\end{equation}%
for any smooth functional $f\in \mathcal{D}(\tilde{\mathcal{G}}^{\ast })$
and arbitrary $Z\in \tilde{\mathcal{G}}^{\ast }.$

It is easy to observe that under the antisymmetry condition $R^{\ast }=-R$
the right-hand side of (\ref{eq2.38}) equals the Lie-Poisson bracket \cite%
{FT,RS-T,RS-T1,Ne,BPS} for the functionals $(X,\bar{T})$ and $(Y,\bar{T})\in
\mathcal{D}(\tilde{\mathcal{G}}^{\ast }).$ Here the adjoint space $\tilde{%
\mathcal{G}}^{\ast }=\tilde{\mathcal{G}}^{\ast }\oplus \mathbb{C}$ is with
respect to a new commutator structure $[\cdot ,\cdot ]_{R}$ on the centrally
extended Lie algebra $\hat{\mathcal{G}}:$ for any $(X,c),(Y,r)\in \hat{%
\mathcal{G}}$ with commutator
\begin{eqnarray}
\lbrack (X,c),(Y,r)]_{R} &:&=\left( [X,Y]_{R},(\frac{d}{dx}X,R(Y))_{-1}+(%
\frac{d}{dx}R(X),Y)_{-1}\right) .  \label{eq2.40} \\
&&  \notag
\end{eqnarray}%
In \ (\ref{eq2.40}) the classical $R$-structure on the Lie algebra $\tilde{%
\mathcal{G}}$ $[X,Y]_{R}:=[R(X),Y]+[X,R(Y)]$ under some conditions on the
mapping $R:\tilde{\mathcal{G}}\rightarrow \tilde{\mathcal{G}}$ \ can
generate on $\tilde{\mathcal{G}}$ a new Lie structure (which it must not).

The above results can be formulated as follows.

\begin{proposition}
\label{P2.1} The Marsden-Weinstein reduced canonical Poisson structure on
the phase space $\bar{M}$ for the monodromy matrix $\bar{T}(\bar{l})\in
\tilde{\mathcal{G}}$ exactly coincides with the corresponding classical
Lie-Poisson AKS-bracket on the centrally extended basis Lie algebra $\hat{%
\mathcal{G}}$ subject to the $R$-structure (\ref{eq2.40}) when it is
antisymmetric.
\end{proposition}

If the antisymmetry property for the mapping $R:\tilde{\mathcal{G}}%
\rightarrow \tilde{\mathcal{G}}$ does not hold, the generated Lie-Poisson
type bracket on the functional space $\mathcal{D}(\tilde{\mathcal{G}}^{\ast
})$ can be, owing to (\ref{eq2.38}), defined as follows: for any $f,g\in
\mathcal{D}(\tilde{\mathcal{G}}^{\ast })$ the bracket
\begin{eqnarray}
\left\{ f(\bar{l}),g(\bar{l})\right\} _{\xi } &:&=(\bar{l},[\nabla f(\bar{l}%
),\nabla g(\bar{l})]_{R})_{-1}+\left( \frac{d}{dx}\nabla f(\bar{l}),R(\nabla
g(\bar{l}))\right) _{-1}+\left( \frac{d}{dx}(R\nabla f(\bar{l})),\nabla g(%
\bar{l})\right) _{-1}  \label{eq2.41} \\
&&  \notag
\end{eqnarray}%
where the generalized $R$-structure $[\cdot ,\cdot ]_{R}$ on $\tilde{%
\mathcal{G}}$ is given by the expression (\ref{eq2.37a}).

\section{$D$-structure and the generalized $R$-structure relationship
analysis.}

As it was stated above, the reduced Poisson bracket on the phase space $\bar{%
M}_{\xi }$ is
\begin{equation}
\left\{ (X,\bar{T}),(Y,\bar{T})\right\} _{\xi }=(\bar{T},[X,Y]_{D})_{-1},
\label{eq3.1}
\end{equation}%
where for any $X,Y\in \tilde{\mathcal{G}}$ the corresponding $D$-structure
on the Lie algebra $\tilde{\mathcal{G}}$ is defined by the classical
expression (\ref{eq2.22}) and the mapping (\ref{eq2.23}). It is natural to
assume that there exists a relationship between the $D$-structure $D:\tilde{%
\mathcal{G}}\rightarrow \tilde{\mathcal{G}}$ and the $R$-structure $R:\tilde{%
\mathcal{G}}\rightarrow \tilde{\mathcal{G}},$ described above in Section 3.

Assume, for brevity, that the $R$-structure (\ref{eq2.36}) is antisymmetric,
that is $R^{\ast }=-R.$ Then it is easy to check that the following
algebraic relationship
\begin{equation}
D(X):=\frac{1}{2}R(\bar{T}X+X\bar{T})  \label{eq3.2}
\end{equation}%
holds for any $X\in \tilde{\mathcal{G}}.$ In fact, the expression (\ref%
{eq2.34}) is equivalent to the following
\begin{equation}
\left\{ (\bar{T},X),(\bar{T},Y)\right\} _{\xi }=(\bar{T}X,R(\bar{T}%
Y))_{-1}-(X\bar{T},R(Y\bar{T}))_{-1}.  \label{eq3.3}
\end{equation}%
Now, substituting the expression (\ref{eq3.2}) into (\ref{eq2.21}), one
obtains that
\begin{eqnarray}
&&\left\{ (\bar{T},X),(\bar{T},Y)\right\} _{\xi }=  \label{3.4} \\
&=&\frac{1}{2}(\bar{T},[R(\bar{T}X+X\bar{T}),Y]+[X,R(\bar{T}Y+Y\bar{T}%
)])_{-1}=  \notag \\
&=&\frac{1}{2}([Y,\bar{T}],R(\bar{T}X))_{-1}+\frac{1}{2}([Y,\bar{T}],R(X\bar{%
T}))_{-1}+  \notag \\
&&+\frac{1}{2}([\bar{T},X],R(\bar{T}Y))_{-1}+\frac{1}{2}([\bar{T}X,R(Y\bar{T}%
)])_{-1}=  \notag \\
&=&\frac{1}{2}(Y\bar{T},R(\bar{T}X))_{-1}-\frac{1}{2}(\bar{T}Y,R(\bar{T}%
X))_{-1}+  \notag \\
&&+\frac{1}{2}(Y\bar{T},R(X\bar{T}))_{-1}-\frac{1}{2}(\bar{T}Y,R(X\bar{T}%
))_{-1}+  \notag \\
&&+\frac{1}{2}(\bar{T}X,R(\bar{T}Y))_{-1}-\frac{1}{2}(X\bar{T},R(\bar{T}%
X))_{-1}+  \notag \\
&&+\frac{1}{2}(\bar{T}X,R(Y\bar{T}))_{-1}-\frac{1}{2}(X\bar{T},R(Y\bar{T}%
))_{-1}=  \notag \\
&=&(\bar{T}X,R(\bar{T}Y))_{-1}-(X\bar{T},R(Y\bar{T}))_{-1},  \notag \\
&&  \notag
\end{eqnarray}%
which coincides exactly with (\ref{eq3.3}).

Rewrite now for convenience the operator relationship (\ref{eq2.28c}) in the
tensor form as
\begin{equation}
(\bar{l}\otimes \mathbb{I})D-D(\mathbb{I}\otimes \bar{l})-\frac{d\ }{dx}D=%
\mathbb{I},  \label{eq3.5}
\end{equation}%
where the tensor $D\in \tilde{\mathcal{G}}\otimes \tilde{\mathcal{G}}^{\ast }
$, owing to the action (\ref{eq3.2}), equals
\begin{equation}
D=\frac{1}{2}(R(\mathbb{I}\otimes \bar{T})+(\mathbb{I}\otimes \bar{T})R).
\label{eq3.6}
\end{equation}%
Substituting the expression (\ref{eq3.6}) into the equation (\ref{eq3.5})
and taking into account the determining equation (\ref{eq2.35})
\begin{equation}
\lbrack \bar{l}\otimes \mathbb{I}+\mathbb{I}\otimes \bar{l},R]-\frac{d}{dx}R=%
\bar{\Omega},  \label{eq3.7}
\end{equation}%
one obtains the relationship for the tensor $\bar{\Omega}\in \tilde{\mathcal{%
G}}\rightarrow \tilde{\mathcal{G}}^{\ast }$:
\begin{equation}
2\mathbb{I}\otimes \mathbb{I}-\bar{\Omega}=[R,\mathbb{I}\otimes \bar{T}\bar{l%
}]+(\mathbb{I}\otimes \bar{T})R(\mathbb{I}\otimes \bar{l})-(\mathbb{I}%
\otimes \bar{l})R(\mathbb{I}\otimes \bar{T}).  \label{eq3.8}
\end{equation}%
The latter makes two $R$- and $D$-structures on the Lie algebra $\tilde{%
\mathcal{G}}$ compatible. Observe that the $D$-structure (\ref{eq3.2}) is
not antisymmetric even though the $R$-structure was assumed to be
antisymmetric. Concerning the $D$-structure determining equation (\ref{eq3.5}%
) one can anticipate that a study of its solutions would describe a set of
nonlinear dynamical systems on the reduced phase space $\bar{M}_{\xi }$
possessing a priori an infinite hierarchy of mutually commuting conservation
laws.

\section{ Example: a fermionic medium and a related quantum exactly solvable
superradiance model}

\subsection{Model description}

We shall demonstrate that the the quantum superradiance properties \cite{AE}
of a generalized model of a one-dimensional many particle charged fermionic
medium, interacting with an external electromagnetic field, can be
completely described by means a quantum Lax type exactly solvable
Hamiltonian system in a specially constructed Fock space.

Consider a Dirac type $N$-particle Hamiltonian operator of a quantum
superradiance model which is expressed as
\begin{equation}
H_{N}:=i\sum_{j=1}^{N}\sigma _{3}^{(j)}\frac{\partial }{\partial x_{j}}%
\otimes \mathbb{I}\mathbf{-}i\beta \mathbb{I}\mathbf{\otimes }\int_{\mathbb{R%
}}dx\varepsilon ^{+}\varepsilon _{x}+\alpha \sum_{j=1}^{N}\sigma
_{1}^{(j)}\otimes \mathcal{E}(x_{j}),  \label{S1.1}
\end{equation}%
where $\sigma _{3}^{(j)},\sigma _{1}^{(j)},j=1,\ldots N,$ are the usual
Pauli matrices, $\alpha \in \mathbb{R}_{+}$ is an interaction constant, $%
0<\beta <1$ is the light speed in the linearly polarized fermionic medium, $%
\mathcal{E}(x):=\left(
\begin{array}{cc}
\varepsilon (x) & 0 \\
0 & \varepsilon ^{+}(x)%
\end{array}%
\right) $ is the one-mode polarization matrix operator at particle location $%
x\in \mathbb{R}$\ with quantized electric field bose-operators $\varepsilon
(x),\varepsilon ^{+}(x):\Phi _{B}\rightarrow \Phi _{B}$ acting in the
corresponding Fock space $\Phi _{B}$ and \ satisfying the commutation
relationships:%
\begin{align}
\lbrack \varepsilon (x),\varepsilon ^{+}(y)]& =\delta (x-y),  \label{S1.2} \\
\lbrack \varepsilon (x),\varepsilon (y)]& =0=[\varepsilon
^{+}(x),\varepsilon ^{+}(y)]  \notag
\end{align}%
for all $x,y\in \mathbb{R}.$ We note that throughout the sequel we employ
units for which the standard constants $\hslash =1=c.$

By construction, the $N$-particle Hamiltonian operator \ (\ref{S1.1}) acts
in the Hilbert space $L_{2}^{(as)}(\mathbb{R}^{N};\mathbb{C}^{2})\otimes\Phi
_{B},$ where $L_{2}^{(as)}(\mathbb{R}^{N};\mathbb{C}^{2})$ denotes the
square-integrable antisymmetric vector functions on $\mathbb{R}^{N},N\in%
\mathbb{Z}_{+}.$ Correspondingly, the Fock space $\Phi_{B}$ allows the
standard representation as the direct sum
\begin{equation}
\Phi_{B}:=\oplus_{n\in\mathbb{Z}_{+}}L_{2}^{(s)}(\mathbb{R}^{n};\mathbb{C}),
\label{S1.3}
\end{equation}
where $L_{2}^{(s)}(\mathbb{R}^{n};\mathbb{C})$ denotes the space of
symmetric square-integrable scalar functions on $\mathbb{R}^{n},n\in\mathbb{Z%
}_{+}.$

Similarly, the corresponding fermionic Fock space
\begin{equation}
\Phi_{F}:=\oplus_{n\in\mathbb{Z}_{+}}L_{2}^{(as)}(\mathbb{R}^{n};\mathbb{C}%
^{2}),  \label{S1.4}
\end{equation}
can be used to represent \cite{BB,Fe,Is,BPS} the Hamiltonian operator (\ref%
{S1.1}) in the second quantized form%
\begin{equation}
\mathbf{H:}=i\int_{\mathbb{R}}dx[\psi_{1}^{+}\psi_{1,x}-\psi_{2}^{+}\psi
_{2,x}-\beta\varepsilon^{+}\varepsilon_{x}+i\alpha(\varepsilon\psi_{2}^{+}%
\psi_{1}+\varepsilon^{+}\psi_{1}^{+}\psi_{2})],  \label{S1.5}
\end{equation}
which acts on the tensored Fock space $\Phi:=\Phi_{F}\otimes\Phi_{B},$ where
$\Phi_{F},$ defined by \ (\ref{S1.4}), can also be representes as
\begin{equation}
\begin{array}{c}
\Phi_{F}:=\oplus_{n\in\mathbb{Z}_{+}}\mathrm{span}\left\{ \int_{\mathbb{R}%
^{n}}dx_{1}dx_{2}...dx_{n}\varphi_{n}^{(m)}(x_{1},x_{2,}...,x_{n})\times%
\right. \\
\quad\quad\qquad\times\Pi_{j=m+1}^{n}\psi_{1}^{+}(x_{j})\Pi_{k=1}^{m}\left.
\psi_{2}^{+}(x_{k})\left. |0\right\rangle :0\leq m\leq
n;\varphi_{n}^{(m)}\in L_{2}^{(as)}(\mathbb{R}^{n};\mathbb{C}^{2})\right\} \
\end{array}
\label{S1.6}
\end{equation}
and $\left. |0\right\rangle \in\Phi_{F}$ is the corresponding vacuum state,
satisfying the determining conditions
\begin{equation}
\psi_{1}(x)\left. |0\right\rangle =0=\psi_{2}(x)\left. |0\right\rangle
,\varepsilon(x)\left. |0\right\rangle =0  \label{S1.7}
\end{equation}
for all $x\in\mathbb{R}.$ The creation and annihilation operators $\psi
_{j}(x),\psi_{k}^{+}(y):\Phi_{F}\rightarrow\Phi_{F},j,k=1,2,$ \ satisfy the
anti-commuting
\begin{align}
\{\psi_{j}(x),\psi_{k}^{+}(y)\} & =\delta_{j,k}\delta(x-y),  \label{S1.8a} \\
\{\psi_{j}(x),\psi_{k}(y)\} & =0=\{\psi_{j}^{+}(x),\psi_{k}^{+}(y)\}  \notag
\end{align}
and commuting
\begin{align}
\lbrack\varepsilon(x),\psi_{j}(y)] & =0=[\varepsilon(x),\psi_{j}^{+}(y)],
\label{S1.8b} \\
\lbrack\varepsilon^{+}(x),\psi_{j}(y)] &
=0=[\varepsilon^{+}(x),\psi_{j}^{+}(y)]  \notag
\end{align}
relationships for all $x,y\in\mathbb{R}$.

As we are interested in proving the exact integrability of our quantum
superradiance model, it is necessary to find the corresponding Lax type
representation of the Hamiltonian system
\begin{align}
d\psi _{1}/dt& =i[\mathbf{H,}\psi _{1}]=\psi _{1,x}+i\alpha \varepsilon
^{+}\psi _{2},  \notag \\
d\psi _{2}/dt& =i[\mathbf{H,}\psi _{2}]=-\psi _{2,x}+i\alpha \varepsilon
\psi _{1},  \label{S1.9} \\
d\varepsilon /dt& =i[\mathbf{H,}\varepsilon ]=-\beta \varepsilon
_{x}+i\alpha \psi _{1}^{+}\psi _{2},  \notag
\end{align}%
generated by the Hamiltonian operator (\ref{S1.5}) as a usual Heisenberg
flow on the quantum operator manifold $\mathbf{M}:=\{(\psi _{1},\psi
_{2},\varepsilon ;\varepsilon ^{+},\psi _{2}^{+},\psi _{1}^{+})\in \mathrm{%
End}\Phi ^{6}\}$. \ \ The dynamical system \ (\ref{S1.9}) possesses also the
following number operators as conservation laws:
\begin{equation}
\mathbf{N}_{F}:=\int_{\mathbb{R}}dx(\psi _{1}^{+}\psi _{1}+\psi _{2}^{+}\psi
_{2}),\mathbf{N}_{B}:=\int_{\mathbb{R}}dx(\varepsilon ^{+}\varepsilon +\psi
_{2}^{+}\psi _{2}),  \label{S1.9a}
\end{equation}%
commuting \ with each other and with the Hamiltonian operator (\ref{S1.5}):%
\begin{equation}
\lbrack \mathbf{N}_{F},\mathbf{N}_{B}]=0,[\mathbf{H,N}_{F}]=0=[\mathbf{H,N}%
_{B}].  \label{S1.9b}
\end{equation}%
The following proposition states that dynamical system \ (\ref{S1.9}) is Lax
type integrable.\bigskip

\begin{proposition}
\label{Prop_S1}The dynamical system \ (\ref{S1.9}) can be linearized by
means of the quantum Lax type spectral problem%
\begin{equation}
df/dx=l(x;\lambda )f,  \label{S1.10}
\end{equation}%
where the operator matrix $l(x;\lambda )\in \mathrm{End}\Phi ^{3\text{ }}$is%
\begin{equation}
l(x;\lambda ):=\left(
\begin{array}{ccc}
-\frac{i\lambda }{3-\beta } & i\xi _{1}\psi _{1} & i\xi _{2}\psi _{2} \\
i\xi _{1}\psi _{1}^{+} & -\frac{i\lambda }{2\beta } & i\xi _{3}\varepsilon
\\
i\xi _{2}\psi _{2}^{+} & i\xi _{3}\varepsilon ^{+} & \frac{i\lambda (3+\beta
)}{2\beta (3-\beta )}%
\end{array}%
\right)  \label{S1.11}
\end{equation}%
for all $x\in \mathbb{R},$ with $\lambda \in \mathbb{C}\ $\ an arbitrary
time-independent spectral parameter, and
\begin{align}
\xi _{1}& :=\xi _{1}(\alpha ,\beta )=-18\alpha \left[ \frac{(9-3\beta
)(\beta +1)}{\beta +3}\right] ^{1/2}\times  \notag \\
& \times \left( \frac{12\beta }{\beta +3}\right) ^{1/2}\frac{\beta +3}{%
2\beta ^{2}+3\beta +3},  \label{S1.12} \\
\xi _{2}& :=\xi _{2}(\alpha ,\beta )=6\alpha (3-3\beta )^{1/2}\left( \frac{%
12\beta }{\beta +3}\right) ^{1/2}\frac{(9-3\beta )(\beta +1)}{(\beta
-1)(2\beta ^{2}+3\beta +3)},  \notag \\
\xi _{3}& :=\xi _{3}(\alpha ,\beta )=72\alpha \beta \frac{(3(1-\beta ))^{1/2}%
}{(\beta -1)(2\beta ^{2}+3\beta +3)}\left[ \frac{(9-3\beta )(\beta +1)}{%
\beta +3}\right] ^{1/2},  \notag
\end{align}%
are constants depending on the interaction parameter $\alpha \in \mathbb{R}%
_{+}$ and the light speed in the polarized fermionic medium $0<\beta <1.$
\end{proposition}

The quantum dynamical system (\ref{S1.9}) may be also regarded as an exactly
solvable approximation of the three-level quantum model studied in \cite{Bo}
subject to its superradiance properties. Concerning the studies of such
superradiance Dicke type one-dimensional models, it is necessary to mention
the work \cite{Ru} in which it was shown that the well-known quantum
Bloch--Maxwell dynamical system
\begin{align}
d\psi _{1}/dt& =i[\mathbf{\hat{H},}\psi _{1}]=i\alpha \varepsilon ^{+}\psi
_{2},  \label{S1.13} \\
d\psi _{2}/dt& =i[\mathbf{\hat{H},}\psi _{2}]=i\alpha \varepsilon \psi _{1},
\notag \\
d\varepsilon /dt& =i[\mathbf{\hat{H},}\varepsilon ]=-\beta \varepsilon
_{x}+i\alpha \psi _{1}^{+}\psi _{2},  \notag
\end{align}%
generated by the the\ reduced quantum Hamiltonian operator
\begin{equation}
\mathbf{\hat{H}:}=-i\int_{\mathbb{R}}dx[\beta \varepsilon ^{+}\varepsilon
_{x}-i\alpha (\varepsilon \psi _{2}^{+}\psi _{1}+\varepsilon ^{+}\psi
_{1}^{+}\psi _{2})]  \label{S1.14}
\end{equation}%
in the strongly degenerate Fock space $\Phi $ is also exactly solvable.
Moreover, it possesses the corresponding Lax type operator whose spectral
problem \cite{FT,No,BPS} is defined in the space $\Phi ^{3}.$ But the
important problem of constructing the stable physical vacuum for the
Hamiltonian (\ref{S1.14}) was on the whole not discussed in \cite{Ru}, and
neither was the problem of studying the related thermodynamics of quantum
excitations over it. More interesting quantum one-dimensional models with
the Hamiltonian similar to (\ref{S1.5}) describing the quantum interaction
of just fermionic particles and only bosonic particles with an external
electromagnetic field were studied, respectively, in \cite{WO} and \cite{Ku}%
. In these investigations, the quantum localized Bethe states were
constructed and analyzed in detail. The corresponding classical version of
the quantum dynamical system (\ref{S1.9}), called the \textit{three-wave
model}, was studied in \cite{No,ZM} and elsewhere.

It is also worth mentioning here that the spectral operator problem (\ref%
{S1.11}) makes sense only if the light speed inside the polarized fermionic
medium is less than the light speed in a vacuum. This is guaranteed by the
dynamical stability of the quantum Hamiltonian system \ (\ref{S1.9})
following from \ the existence of an additional infinite hierarchy of
conservation laws, suitably determined on the quantum operator phase space $%
\mathbf{M}.$ Consequently, one can expect that the quantum dynamical system (%
\ref{S1.9}) also possesses the many-particle localized photonic states in
the Fock space $\Phi ,$ which are called \textit{quantum solitons}, whose
spatial range is inverse to the number of interior particles, and which can
be interpreted as special Dicke type superradiance laser impulses. In
particular, the quantum stability, solitonic formation aspects and
construction of the physical ground state related with the unbounded \textit{%
a priori} from below Hamiltonian operator (\ref{S1.5}) are of great
importance for physical applications.

In the next subsection we will make use of use of the quantum spectral
problem (\ref{S1.11}) to prove that the quantum dynamical system (\ref{S1.9}%
) allows the standard $R$-matrix description, which makes it possible to
construct an infinite hierarchy of commuting conservation laws, thereby
ensuring its complete quantum integrability.

\subsection{The quantum $R$-matrix structure}

The spectral problem (\ref{S1.11}) belongs to exactly the same class whose
Lie-algebraic properties were studied in Section\ 4. That means, in
particular, that the system of dynamical equations
\begin{align}
d\psi _{1}/dt& =\psi _{1,x}+i\alpha \varepsilon ^{+}\psi _{2},  \notag \\
d\psi _{2}/dt& =-\psi _{2,x}+i\alpha \varepsilon \psi _{1},  \label{S2} \\
d\varepsilon /dt& =-\beta \varepsilon _{x}+i\alpha \psi _{1}^{+}\psi _{2}
\notag
\end{align}%
jointly with their adjoint flows determine on an infinite-dimensional
functional manifold $M\subset C^{\infty }(\mathbb{R}/2\pi \mathbb{Z};\mathbb{%
C}^{6})$ a completely Lax type integrable dynamical system and whose
classical linear spectral \ problem (\ref{S1.10}) entails the corresponding
Poissonian relationships \ (\ref{eq2.34}) between the respectively defined
monodromy matrix \ $T(x;\lambda )\in \mathrm{End}$ $\mathbb{C}^{3}\ $for any
$x\in $ $\mathbb{R}$ and $\lambda \in \mathbb{C}.$ Taking into account the
standard quantization rules \cite{Fa,Sk} one can easily generalize these
Poissonian relationships to the correspondingly defined operator
relationship
\begin{equation}
\mathcal{R(\lambda },\mu )T(x;\lambda )\otimes \mathbb{I}=\mathbb{I\otimes }%
T(x;\mu )\mathcal{R(\lambda },\mu )\   \label{S2a}
\end{equation}%
for some scalar $\mathcal{R}$-matrix $\mathcal{R(\lambda },\mu )\in \mathbb{C%
}^{3}\otimes \mathbb{C}^{3}$ between quantum monodromy operators $%
T(x;\lambda )$ and $T(x;\mu )$ $\in \mathrm{End}\ \Phi ^{3},$ acting already
in the Fock space $\mathrm{\ }\ \Phi ^{3}.$ To realize this scheme, we first
consider the following generalized quantum operator Cauchy problem for the
spectral equation (\ref{S1.10}) subject to the periodic conditions $l(x+2\pi
;\lambda )=l(x;\lambda )\in \mathrm{End}\ \Phi ^{3}$ for all $x\in $ $%
\mathbb{R}$ and $\lambda \in \mathbb{C}:$%
\begin{equation}
dF(x,y;\lambda )/dx=\vdots l(x;\lambda )F(x,y;\lambda )\vdots ,  \label{S3}
\end{equation}%
where $F(x,y;\lambda )\in \mathrm{End}\ \Phi ^{3}$ is the corresponding
fundamental transition operator matrix satisfying
\begin{equation}
\left. F(x,y;\lambda )\right\vert _{y=x}=\mathbb{I},  \label{S3.2}
\end{equation}%
and the operation $\vdots \cdot \vdots $ arranges operators $\psi _{j,}\psi
_{j}^{+},j=\overline{1,2},$ $\varepsilon $ and $\varepsilon ^{+},$ via the
standard normal ordering \cite{Sk,BB} that does not change the position of
any other operators; for instance, $\vdots A\psi _{1}^{+}\psi
_{2}\varepsilon ^{+}B\vdots $\ $=\psi _{1}^{+}\varepsilon ^{+}AB\psi _{2}$
for any $A,B\in \mathrm{End}\ \Phi .$

Construct now the operator products
\begin{equation}
\mathcal{\hat{F}}(x,y|\lambda ,\mu ):=\tilde{F}(x,y;\lambda )\overset{%
\thickapprox }{F}(x,y;\mu ),  \label{S3.3}
\end{equation}%
\begin{equation*}
\mathcal{\check{F}}(x,y|\lambda ,\mu ):=\overset{\thickapprox }{F}(x,y;\mu )%
\tilde{F}(x,y;\lambda ),
\end{equation*}%
where
\begin{align}
\tilde{F}(x,y;\lambda )& :=F(x,y;\lambda )\otimes \mathbb{I}  \label{S3.4} \\
\overset{\thickapprox }{F}(x,y;\mu )& :=\mathbb{I}\otimes F(x,y;\mu )  \notag
\end{align}%
are for all $x,y\in \mathbb{R}$ $,$ $\lambda ,\mu \in \mathbb{C},$ the
corresponding tensor products of operators acting in the space $\Phi
^{3}\otimes \Phi ^{3}.$ The following proposition is crucial \cite%
{Sk,BPS,MBPS} for the further analysis of integrability of\textbf{\ }the%
\textbf{\ }quantum dynamical system (\ref{S1.9}) and is proved by a direct
computation.\medskip

\begin{proposition}
\noindent \label{Prop_S2} \textit{The operator expressions (\ref{S3.3})
satisfy the following differential relationships:
\begin{align}
\frac{\partial }{\partial x}\mathcal{\hat{F}}(x,y|\lambda ,\mu )& =\vdots
\mathcal{\hat{L}(}x;\lambda ,\mu )\mathcal{\hat{F}}(x,y|\lambda ,\mu )\vdots
,  \label{S3.5} \\
\frac{\partial }{\partial x}\mathcal{\check{F}}(x,y|\lambda ,\mu )& =\vdots
\overset{\smallsmile }{\mathcal{L}}\mathcal{(}x;\lambda ,\mu )\mathcal{%
\check{F}}(x,y|\lambda ,\mu )\vdots ,  \notag
\end{align}%
where the matrices%
\begin{align}
\mathcal{\hat{L}(}x;\lambda ,\mu )& =\tilde{l}(\text{ }x;\lambda )+\overset{%
\approx }{l}(x;\mu )-\alpha \hat{\bigtriangleup}(x;\lambda ,\mu ),
\label{S3.6} \\
\overset{\smallsmile }{\mathcal{L}}\mathcal{(}x;\lambda ,\mu )& =\tilde{l}(%
\text{ }x;\lambda )+\overset{\approx }{l}(x;\mu )-\alpha \check{%
\bigtriangleup}(x;\lambda ,\mu ),  \notag
\end{align}%
and $\hat{\bigtriangleup}(x;\lambda ,\mu ),\check{\bigtriangleup}(x;\lambda
,\mu )$ satisfy the algebraic relationship $P\hat{\bigtriangleup}(x;\lambda
,\mu )P=\check{\bigtriangleup}(x;\lambda ,\mu )$ \ for all $x\in \mathbb{R},$
$\lambda ,\mu \in \mathbb{C},$ where $P\in \mathrm{End}\ \Phi ^{3}\otimes
\Phi ^{3}$\ is the standard transmutation operator in the space $\Phi
^{3}\otimes \Phi ^{3},$ that is $P(a\otimes b):=b\otimes a$ for any vectors $%
a,b\in \Phi ^{3}.$}
\end{proposition}

Using Proposition \ref{Prop_S1} one can easily verify that there exists a
scalar $\mathcal{R}$-matrix $\mathcal{R(\lambda },\mathcal{\mu )},$ $%
\mathcal{R\in }\mathrm{End}\mathbb{C}^{9},$ such that
\begin{equation}
\mathcal{R(\lambda },\mu )\mathcal{\hat{L}(}x;\lambda ,\mu )=\overset{%
\smallsmile }{\mathcal{L}}\mathcal{(}x;\lambda ,\mu )\mathcal{R(\lambda }%
,\mu )  \label{S3.7}
\end{equation}%
holds for all $\lambda ,\mu \in \mathbb{C}$ and $x\in \mathbb{R}.$ This,
owing to the equations \ (\ref{S3.5}), implies the main functional
Yang--Baxter type \cite{Fa,Sk,Ku,MBPS} operator relationship%
\begin{equation}
\mathcal{R(\lambda },\mu )\mathcal{\hat{F}}(x,y|\lambda ,\mu )=\mathcal{%
\check{F}}(x,y|\lambda ,\mu )\mathcal{R(\lambda },\mu )  \label{S3.8}
\end{equation}%
is satisfied for any $x,y\in \mathbb{R}$ and $\lambda ,\mu \in \mathbb{C},$
where
\begin{equation}
\mathcal{R(\lambda },\mu )=(\lambda -\mu )P-i\alpha \mathbb{I}\mathbf{\ }
\label{S3.9}
\end{equation}%
is the corresponding the quantum $\mathcal{R}$-operator. Recalling now that
periodicity condition, from (\ref{S3.8}) one easily deduces by means of the
trace-operation that the monodromy operator matrix $T(x;\lambda ):=F(x+2\pi
,x;\lambda )$ satisfies the algebraic expression
\begin{equation}
\mathcal{R(\lambda },\mu )T(x;\lambda )\otimes \mathbb{I}=\mathbb{I\otimes }%
T(x;\mu )\mathcal{R(\lambda },\mu ),  \label{s3.9a}
\end{equation}%
giving rise, for all $x\in $ $\mathbb{R}$ and $\lambda ,\mu \in $ $\mathbb{C}%
,\ \ $to the following commutation relationship
\begin{equation}
\left[ \mathrm{tr}T(x;\lambda ),\mathrm{tr}T(x;\mu )\right] =0.
\label{S3.11}
\end{equation}%
Actually, it follows from (\ref{S3.8}) that
\begin{align}
\mathrm{tr}(T(x;\lambda )\otimes T(x;\mu ))& =\mathrm{tr}(\mathcal{R}%
^{-1}T(x;\mu )\otimes T(x;\lambda ))=  \label{S3.12} \\
& =\mathrm{tr}(T(x;\mu )\otimes T(x;\lambda )).  \notag
\end{align}%
Taking into account that $\mathrm{tr}(A\otimes B)=\mathrm{tr}A\cdot \mathrm{%
tr}B$ for any operators $A,B\in \mathrm{End}\ \Phi ^{3},$ one easily obtains
(\ref{S3.11}) from (\ref{S3.12}). Consequently, the $\lambda -$dependent
operator functional
\begin{equation}
\gamma (\lambda ):=\mathrm{tr}T(x,\lambda )\cong \sum_{j\in \mathbb{Z}%
_{+}}\gamma _{j}\lambda ^{-j},  \label{S3.13}
\end{equation}%
as $\left\vert \lambda \right\vert \rightarrow \infty $ generates an
infinite hierarchy of commuting conservation laws $\gamma _{j}:\Phi
\rightarrow \Phi ,j\in \mathbb{Z}_{+}$ $:$%
\begin{equation}
\left[ \gamma _{j},\gamma _{k}\right] =0  \label{S3.14}
\end{equation}%
for all $j,k\in \mathbb{Z}_{+},$ where, in particular,
\begin{align}
\gamma _{1}& =\mathbf{N}_{F}=\int_{\mathbb{R}}dx(\psi _{1}^{+}\psi _{1}+\psi
_{2}^{+}\psi _{2}),\text{ \ }\gamma _{2}=\mathbf{N}_{B}=\int_{\mathbb{R}%
}dx(\varepsilon ^{+}\varepsilon +\psi _{2}^{+}\psi _{2}),  \label{S3.15} \\
\gamma _{3}& =\mathbf{P}=i\int_{\mathbb{R}}dx(\psi _{1}^{+}\psi _{1,x}+\psi
_{2}^{+}\psi _{2,x}+\varepsilon ^{+}\varepsilon _{x}),\text{ }  \notag \\
\gamma _{4}& =\mathbf{H}=i\int_{\mathbb{R}}dx\left[ \psi _{1}^{+}\psi
_{1,x}-\psi _{2}^{+}\psi _{2,x}-\varepsilon ^{+}\varepsilon _{x}+i\alpha
(\varepsilon \psi _{2}^{+}\psi _{1}+\psi _{1}^{+}\psi _{2}\varepsilon ^{+})%
\right] .  \notag
\end{align}%
Since the operator functional $\gamma _{4}=\mathbf{H}$ is the Hamiltonian
operator for the dynamical system (\ref{S1.9}), from (\ref{S3.14}) one
obtains
\begin{equation}
\left[ \mathbf{H},\gamma _{j}\right] =0  \label{S3.16}
\end{equation}%
for all $j\in $ $\mathbb{Z}_{+}$; that is, all of functionals $\gamma
_{j}:\Phi \rightarrow \Phi ,j\in \mathbb{Z}_{+},$ are conservation laws.

Moreover, making use of the exact operator relationships (\ref{S3.8}) one
can easily construct the physically stable quantum states $|(N,M)>\in \Phi $
for all $N,M\in \mathbb{Z}_{+}$ upon redefining the Fock vacuum $\left.
|0\right\rangle \in \Phi ,$ which is nonphysical for the dynamical system (%
\ref{S1.9}), governed by the unbounded from below Hamiltonian operator (\ref%
{S1.5}). Following a renormalization scheme similar to those developed in
\cite{Fa,YK,MBPS}, one can construct a new physically stable vacuum
\begin{equation}
\left. |(0)\right\rangle _{phys}:=\prod\limits_{q\leq \mu _{j}\leq
Q}B^{+}(\mu _{j})\left. |0\right\rangle  \label{S3.17}
\end{equation}%
by means of the new commuting to each other "creation" operators $B^{+}(\mu
):\Phi \rightarrow \Phi ,\mu \in \mathbb{C},$ generated by suitable
components of the monodromy operator matrix $T(x;\mu ):\Phi ^{3}\rightarrow
\Phi ^{3},$ $x\in \mathbb{R},$ whose commutation relationships with the
Hamiltonian operator (\ref{S1.5})
\begin{equation}
\lbrack \mathbf{H,}B^{+}(\mu )]=S(\mu ;\alpha ,\beta )B^{+}(\mu )
\label{S3.18}
\end{equation}%
are parameterized by the two-particle scalar scattering factor $S(\mu
;\alpha ,\beta ),\mu \in \mathbb{C},$ and where values $q<Q\in \mathbb{R}$
are to be determined \cite{Fa} from the condition that quantum excitations
over the physical vacuum \ (\ref{S3.17}) have positive energy. Since the
physical vacuum (\ref{S3.17}) is an eigenstate of the Hamiltonian operator (%
\ref{S1.5}), the corresponding quantum eigenstates of the excitations can be
represented as
\begin{equation}
\left. |(\mu )\right\rangle :=B^{+}(\mu )\left. |(0)\right\rangle _{phys}
\label{S3.19}
\end{equation}%
for some $\mu \in \mathbb{R}$ and the new energy level can be taken into
account in the renormalized Hamiltonian operator (\ref{S1.5}) by means of
the chemical potentials $a_{F},a_{B}\in \mathbb{R}:$%
\begin{equation}
\mathbf{H}_{a}:=\mathbf{H}-a_{F}\mathbf{N}_{F}-a_{B}\mathbf{N}_{B},
\label{S3.20}
\end{equation}%
which should be determined from the conditions
\begin{equation}
\mathbf{H}_{a}\left. |(0)\right\rangle _{phys}=0,\;\left\langle (\mu
)|\right. \mathbf{H}_{a}\left. |(\mu )\right\rangle >0  \label{S3.21}
\end{equation}%
for any $\mu \in \mathbb{R}.$ The physical vacuum state and quantum
Hamiltonian renormalization construction described above make it possible to
study the properties of superradiance quantum photonic impulse structures
generated by interaction of the charged fermionic medium with an external
electromagnetic field. Owing to the existence of quantum periodic
eigenstates over the physically stable vacuum, one can also investigate the
related thermodynamic properties of the model and analyze the generated
superradiance photonic structures, which are important for explaining many
\cite{AE} existing experiments.

\section{Conclusion}

We have considered the standard canonically symplectic phase space $%
M:=T^{\ast }(\tilde{\mathcal{G}})$, generated by the centrally extended
basis manifold to be an affine loop Lie algebra $\tilde{\mathcal{G}}$ on the
circle $\mathbb{S}^{1}.$ Subject to the standard Hamiltonian Lie algebra $%
\tilde{\mathcal{G}}$-action on $M,$ with respect which the symplectic
structure on $M$ is invariant, constructed the corresponding momentum
mapping and carried out the standard Marsden-Weinstein reduction of the
manifold $M$ upon the reduced phase space $\bar{M}_{\xi }$ endowed with the
reduced Poisson bracket $\left\{ \cdot ,\cdot \right\} _{\xi }$. The latter
allows to construct on the phase space $\bar{M}_{\xi }$ commuting to each
other vector fields which are equivalent to some nonlinear dynamical systems
possessing an infinite hierarchy of commuting conservation laws. Moreover,
these mentioned commuting vector fields on $\bar{M}_{\xi }$ realize exactly
their corresponding Lax type representations.

Presented detailed analysis of commutation properties for the related flows
on the basis manifold makes it possible to define a suitable $D$-structure
on the Lie algebra $\tilde{\mathcal{G}}$, deeply related with the
corresponding classical $R$-structure on $\tilde{\mathcal{G}}$, generated by
the reduced Poisson bracket on the phase space $\bar{M}_{\xi }$. As a
bi-product of our analysis we stated that these $R$- and $D$-structures are
completely equivalent to a suitably generalized classical
Lie-Poisson-Adler-Kostant-Symes-Kirillov-Berezin structure on the adjoint
space $\hat{\mathcal{G}}^{\ast }$. We derived also the determining equation
for the $D$-structure, classifying the generalized Lax type integrable
nonlinear dynamical systems on the reduced phase space $\bar{M}_{\xi }$,
whose respectively defined $R$-structures are not necessary both
antisymmetric and local, as it was before described in \cite{ArMe,ABT} by
means of an other approach. It is worth also to mention that the reduction
scheme devised in this work can be applied also to the centrally extended
algebra of pseudo-differential operators and affine loop algebras on the
circle $\mathbb{S}^{1}.$

As an example of the physical Hamiltonian system whose exact solvability can
be stated by means of a related $R$-structure, we have proposed a new
generalized superradiance model, describing a one-dimensional many particle
charged fermionic medium interacting with an external electromagnetic field.
Its operator structure allows to calculate by means of the $R$-matrix
approach diverse superradiance effects, which are closely related to the
formation of the bound quantum solitonic states and their stability. The
existence of these states is established by suitable applying the physical
vacuum renormalization subject to which all quantum excitations are of
positive energy. This procedure, based on the determining operator
relationships (\ref{S3.8}), enables one to describe the thermodynamic
properties of the quantum dynamical system over the stable physical vacuum.
In addition, it facilitates analysis of the corresponding thermodynamic
states of the resulting quantum photonic system and its superradiance
properties. Our work indicates that a more detailed investigation of these
and related topics is in order, which we plan to undertake elsewhere.

\section*{Acknowledgements}

Authors are cordially thankful to Prof. D. Blackmore (NJIT, NJ USA) for
interest in the work, fruitful discussions and valuable remarks.


\begin{thebibliography}{99}
\bibitem{AM} Abraham R., Marsden J.E. Foundations of mechanics.
Benjamin/Cummins Publisher, (1978)

\bibitem{AE} Allen A. Eberley J.H. Optical resonance and two-level atoms.
Wiley, 1975

\bibitem{Ar} Arnold V.I. Mathematical methods of classical mechanics.
Springer (1989)

\bibitem{ArMe} Arutyunov G.E., Medvedev P.B. Generating equation for $r$%
-matrices related to the dynamical systems of Calogero type. Phys. Lett.,
A223, (1996), pp. 66-74

\bibitem{ABT} Avan J., Babelon O., Talon M. Construction of classical $R$%
-matrices for the Toda and Calogero models. Alg. Anal. 6(2) (1994) p.67

\bibitem{BV} Babelon O., Viallet C-M. Hamiltonian structures and Lax
equations. Phys.Lett. B, 237(3,4) (1990), p. 411-416

\bibitem{BPS} Blackmore D., Prykarpatsky A.K., Samoylenko V.H. Nonlinear
dynamical systems of mathematical physics. World Scientific Publisher,
(2011) p. 542

\bibitem{Bo} Bogolubov (Jr.) et al. JINR Rapid Communications, 3, 1984, p. 26

\bibitem{BB} Bogolubov N.N, Bogolubov N.N. (jr.) Introduction to quantum
statistical mechanics. World Scientific, Second Edition, 2009

\bibitem{CD} Calogero F. and Degasperis A. Spectral Transform and Solitons,
v.1, North-Holland, Amsterdam, 1982

\bibitem{Fa} Faddeev L.D. Quantum completely integrable field theories. In
Proceedings of the "Fifth Alushta Conference on nonlocal field theories",
Dubna, 1979, p. 249-299 (in Russian)

\bibitem{FT} Faddeev L.D., Takhtadjan L.A Hamiltonian methods in the theory
of solitons. Springer, (2000)

\bibitem{Fe} Feynman R.P. Statistical mechanics. W.A. Benjamin, Inc. 1972

\bibitem{Is} Ishihara A. Statistical physics, Academic Press, New York, 1971

\bibitem{Ko} Korepin V.E. Direct calculation of the S-matrix in the massive
Thirring model. Theor. Math. Physics, 41(2), 1979, p. 169-189 (in Russian)

\bibitem{Ku} Kulish P.P. Quantum nonlinear wave interaction model. Physica,
D12, 1986, p. 360-364

\bibitem{MBPS} Mitropolski Yu.A., Bogolubov N.N. (jr.), Prykarpatsky A.K.
and Samoylenko V.H. Integrable dynamical systems: differential-geometric and
spectral aspects. K.: "Naukova Dumka", 1987

\bibitem{Ne} Newell A. Solitons in mathematics and physics. SIAM, 1985

\bibitem{No} Novikov S.P. (Editor) Theory of solitons. Springer, (1984)

\bibitem{Oh} Ohkuma K. Thermodynamics of the quantum three wave interaction
model. Journal Phys. Soc. Japan, 54(8), 1985, p. 2817-2828

\bibitem{PSP} Prykarpatsky Y.A., Samoilenko A.M., Prykarpatsky A.K. The
geometric properties of canonically reduced symplectic spaces with symmetry,
their relationship with structures on associated principal fiber bundles and
some applications. Opuscula Mathematica, 25(2) (2005), p 287-298

\bibitem{PM} Prykarpatsky A., Mykytyuk I. Algebraic integrability of
nonlinear dynamical systems onmanifolds: classical and quantum aspects.
Kluwer Academic Publishers, the Netherlands, (1998)

\bibitem{RS-T1} Reyman A.G., Semenov-Tian-Shansky M.A. Integrable Systems,
The Computer Research Institute Publ., Moscow-Izhvek, 2003 (in Russian)

\bibitem{RS-T2} Reyman A.G. and Semenov-Tian-Shansky M.A. The Hamiltonian
structure of Kadomtsev-Petviashvili type equations, LOMI Proceedings, Nauka,
Leningrad, 164, (1987) p. 212-227 (in Russian)

\bibitem{RS-T} Reyman A.G. and Semenov-Tyan-Shansky M.A. "Reduction of
Hamiltonian systems, affine Lie algebras, and Lax equations, I, II," Invent.
Math., 54, No. 1,(1979), p. 81-100, and 63, No. 3, (1981), p. 423-432

\bibitem{Ru} Rupasov V.I. Contribution to the Dicke superradiance theory.
Exact solution of the quasi-onedimensional quantum model. Sov. Phys. JETP,
56(5), 1982, p. 989-995

\bibitem{S-T} Semenov-Tian-Shansky M.A. What is a R-matrix. Functional
analysis and its applications. Vol. 17, No. 4, (1983), p. 259-272

\bibitem{Sk} Sklyanin E.K. Quantum variant of the inverse scattering
transform method. Proceedings of LOMI, 1980 (95) pp. 55-128 (in Russian)

\bibitem{Ts} Tsyplyaev S.A. Commutation relations for transition matrix in
classical and quantum inverse scatering method. Theor. Math.Phys., 48(1)
(1981) pp. 24-33 (in Russian)

\bibitem{WO} Wadati M. and Ohkuma K. Bethe states for the quantum three wave
interaction equation. Journal Phys. Soc. Japan, 53(4),1984, p. 1229-1237

\bibitem{YK} Yukhnovsky I.R., Kozlovsky M.P. and Pylyuk I.V. Microscopic
phase transitions theory in three-dimensional systems. Lviv, Eurosvit
Publisher, 2001

\bibitem{ZM} Zakharov V.E. and Manakov S.V. JETP, 42, 1976, p. 842
\end{thebibliography}
\end{document}